\documentclass[a4paper,11pt]{article}
\pdfoutput=1 % if your are submitting a pdflatex (i.e. if you have
\usepackage{jcappub} % for details on the use of the package, please
\usepackage{graphicx}% Include figure files
\usepackage{dcolumn}% Align table columns on decimal point
\usepackage{bm}% bold math
\usepackage{csquotes}
\usepackage{xcolor}
\usepackage{hyperref}% add hypertext capabilities  
\usepackage{subcaption}
\usepackage{enumitem}

\usepackage[T1]{fontenc} % if needed

\title{Axions and Primordial Magnetogenesis: the Role of Initial Axion Inhomogeneities}

\author[a,b]{Filippo Anzuini }
\author[c]{, Angelo Maggi}

\affiliation[a]{School of Physics, The University of Melbourne, Parkville, Victoria 3010, Australia}
\affiliation[b]{Australian Research Council Centre of Excellence for Gravitational Wave Discovery (OzGrav), The University 
of Melbourne, Parkville, Victoria 3010, Australia}
\affiliation[c]{Theoretical Particle Physics and Cosmology, King’s College London, Strand, London WC2R 2LS, UK}

\emailAdd{filippo.anzuini@gmail.com}
\emailAdd{angelo.maggi@kcl.ac.uk}

\abstract{The relic density of dark matter in the $\Lambda$CDM model restricts the parameter space for a cosmological axion field, constraining the axion decay constant, the initial amplitude of the axion field and the axion mass. It is shown via lattice simulations how the relic density of axion-like particles with masses close to the one of the QCD axion is affected by axion-gauge field interactions and by initial axion inhomogeneities. For pre-inflationary axions, once the Hubble parameter becomes smaller than the axion mass, the latter starts to oscillate, and part of its energy density is spent producing gauge fields via parametric resonance. If the gauge fields are dark photons and Standard Model photons, the energy density of dark photons becomes higher than the one of the axion, while the high conductivity of the primordial plasma damps the oscillations of the photon field. Such a scenario allows for the production of small-scale, primordial magnetic fields, and it is found that the relic density of axions with a low decay constant are within the bounds set by the $\Lambda$CDM model, while GUT-scale axions are far too abundant. It is also shown that initial inhomogeneities of the axion field can change substantially the gauge field production, boosting or suppressing (depending on the axion parameters and couplings) the magnetogenesis mechanism with respect to an homogeneous axion field. It is found that when the axion mass is far lighter than the QCD axion model and the initial axion field is inhomogeneous, weak but cosmologically relevant magnetic field seeds can be generated on scales of the order of $0.1$ kpc.
}

\begin{document}
\maketitle
\flushbottom

\section{Introduction}
\label{sec:intro}

Axions were first introduced as a dynamical solution to the strong-CP problem (QCD axions) \citep{Peccei_1977a, Peccei_1977b, Wilczek_1978, Weinberg_1978, Cortona_2016, Di_Luzio_2020}. They are pseudo Nambu-Goldstone bosons emerging from the breaking of the $U(1)$ Peccei-Quinn symmetry, with masses and couplings to Standard Model particles which follow a strict, inverse proportionality to the energy scale at which the Peccei-Quinn symmetry is broken. However, several theoretical and experimental efforts \citep{Irastorza_2018} are devoted to the search of light and ultra-light generalizations of QCD axions (generally called axion-like particles, or ALPs), whose masses and couplings differ from the QCD axion model. From a cosmological point of view, ALPs have been proposed as candidates for the inflaton field driving the cosmic expansion during inflation \citep{Freese_1990, Pajer_2013}, they may constitute the totality of dark matter \citep{Marsh_2016}, and lead to a rich phenomenology when coupled to other Beyond Standard Model and Standard Model particles \citep{Kitajima_2018, Agrawal_2018}, such as magnetogenesis \citep{Choi_2018} and the production of primordial gravitational waves \citep{Barnaby_2012, Dimastrogiovanni_2017, Ratzinger_2021, Kitajima_2021}. 

Current constraints on the relic density of dark matter \cite{Planck_2020} restrict the viable parameter space for axions, if they constitute the entirety of dark matter. This translates on bounds on the axion decay constant $f$, the initial amplitude of the axion field and its mass. For the QCD axion, the decay constant is constrained to $f \lesssim 10^{12}$ GeV (for an initial amplitude of the axion field of order $\approx f$), while higher values of $f$ are compatible with the $\Lambda$CDM model only if the initial amplitude of the axion field is $\ll f$. One way to relax such constraints and deplete the axion abundance is allowing for the coupling of axions (either QCD axions or ALPs) with gauge fields in the Early Universe, leading to the sourcing of $U(1)$ Abelian gauge fields (such as dark photons) \citep{Kitajima_2018, Agrawal_2018, Choi_2018}, or $SU(2)$ non-Abelian gauge fields (see for example \citep{Domcke_2019}, where the axion plays the role of the inflaton field). If the axion-gauge field couplings are sufficiently large, axion oscillations lead to a resonant production of the gauge fields \citep{Kofman_1997, Amin_2015, Agrawal_2018, Kitajima_2018}, which terminates when the gauge field abundance becomes comparable to the one of the axion, while the latter suffers a sharp drop due to the backreaction of the sourced gauge fields. For the specific case of the axion-photon and axion-photon-dark photon coupling, axions may lead to the generation of primordial magnetic fields. The magnetic fields produced in the primordial universe may constitute the seeds for the large-scale, feeble magnetic fields that permeate galaxies, galaxy clusters and cosmic voids \citep{Durrer_2013, Tashiro_2013, Choi_2018}.

In this work we focus on the interaction of ALPs with Abelian gauge fields, and investigate via lattice simulations how the production of primordial, standard and dark electromagnetic fields at the expense of the axion field leads to a depletion of the axion abundance.  We first simulate axions with decay constants above the inflationary energy scale (with $f \in [10^{14}, 10^{16}]$ GeV), and assume that the Peccei-Quinn symmetry is not restored after the end of inflation, i.e. that the reheating temperature $T_R$ satisfies $T_R < f$. In these conditions the axion field is initially homogeneous, and we find that while the ALP-photon coupling does not lead to a significant magnetic field production, large couplings among axions, dark photons and photons \citep{Choi_2020, Hook_2023} can overcome the damping of photon oscillations due to the high conductivity of the primordial plasma \citep{Baym_1997}, producing early seeds of small-scale primordial magnetic fields. Such a mechanism was exploited in \citep{Choi_2018} for ultra-light ALPs after the Big Bang Nucleosynthesis epoch. Here we extend those results and apply them to earlier times, when the temperature of the universe $T$ is in the range $T \gtrsim 1$ MeV, and to heavier ALPs (with masses in the range $10^{-10}$-$10^{-8}$ eV) compared to the case studied in \citep{Choi_2018}, where masses of the order of $10^{-17}$ eV were considered. We find that while axions with $f \lesssim 10^{15}$ GeV have an abundance lower than $\Omega_{DM}$ (where $\Omega_{DM}$ is the dark matter relic density \citep{Planck_2020}), axions with $f$ close to the GUT scale remain overabundant. We also perform simulations where the axion field is initially inhomogeneous, as it would be the case for example for axion field configurations emerging from the decay of a network of strings. In particular, we show that in the latter case the magnetogenesis mechanism can be far less efficient than in the pre-inflationary scenario. However, for certain coupling strengths and axion parameters, the intensity of the sourced magnetic fields increases with respect to scenarios where magnetogenesis is triggered by an initially homogeneous axion field. We also show that ultra-light ALPs can produce weak, standard magnetic fields on scales of the order of $0.1$ kpc.

This work is structured as follows. In Section \ref{sec:dynamics} we outline the model lagrangian density employed throughout this work. We also describe how initial inhomogeneities in the axion field can arise. The results of our lattice simulations are presented in Section \ref{sec:lattice_sim}.

\section{Axion and gauge field dynamics}
\label{sec:dynamics}
In this section we discuss the microphysical model that we input in our lattice simulations. We first describe in Section \ref{sec:photons} the prototype model and introduce the relevant couplings, assuming that the axion field interacts with dark photons and photons. We highlight the role played by the primordial plasma and of the source terms appearing in the equations of motion. We then turn to inhomogeneities in the initial axion field configuration in Section \ref{sec:Fluctuations}.

\subsection{Homogeneous axion field}
\label{sec:photons}
The lagrangian density for a system composed by ALPs, dark photons and photons reads \citep{Kitajima_2018, Choi_2018}

\begin{align}
\mathcal{L} & = \frac{1}{2}\partial_\mu\phi\partial^\mu \phi - V(\phi)  -\frac{1}{4}F_{\mu\nu}F^{\mu\nu} - \frac{g_{\phi A}}{4f}\phi F_{\mu\nu}\tilde{F}^{\mu\nu} - \frac{1}{4}X_{\mu\nu}X^{\mu\nu}  \nonumber \\ 
& - \frac{g_{\phi X}}{4f}\phi X_{\mu\nu}\tilde{X}^{\mu\nu} - \frac{g_{\phi A X}}{2f}\phi F_{\mu\nu}\tilde{X}^{\mu\nu}  + J^{\mu}A_\mu \ .
\label{eq:lagrangian_density}
\end{align}

In Eq. \eqref{eq:lagrangian_density}, $\phi$ is the ALP field, and the ALP potential reads
\begin{equation}
V(\phi) = m^2_\phi f^2\bigg[1-\cos\bigg(\frac{\phi}{f}\bigg)\bigg] \ ,
\end{equation}
where $m_\phi$ is the ALP mass. The field strength tensor $F_{\mu\nu}$ is given by $F_{\mu\nu} = \partial_{\mu}A_{\nu} - \partial_{\nu}A_{\mu}$ (where $A_\mu$ is the photon field), while $X_{\mu\nu} = \partial_{\mu}X_{\nu} - \partial_{\nu}X_{\mu}$ is the corresponding one for the dark photon field ($X_\mu$). The $\tilde{F}_{\mu\nu}$ and $\tilde{X}_{\mu\nu}$ tensors correspond to the duals of $F_{\mu\nu}$ and $X_{\mu\nu}$ respectively. The coupling constants for the axion-photon, axion-dark photon and axion-photon-dark photon interactions are denoted by $g_{\phi A}, g_{\phi X}$ and $g_{\phi A X}$ respectively \cite{Agrawal_2018, Kitajima_2018, Choi_2018, Choi_2020, Hook_2023}, which in principle can be large \citep{Plakkot_2021}, while $J^\mu$ is the standard four-dimensional electromagnetic current. Note that in Eq. \eqref{eq:lagrangian_density} we neglect a possible kinetic mixing between dark photons and photons, i.e. a lagrangian density contribution of the form $\epsilon F_{\mu\nu}X^{\mu\nu}$ \cite{Fabbrichesi_2020}. Due to the smallness of $\epsilon$ (typically $\epsilon \lesssim 10^{-3}$), we expect a negligible correction with respect to the other interaction terms proportional to the spacetime gradients of the axion field. For simplicity we will consider the kinetic mixing term in a separate work.

Using the FRW background metric 

\begin{equation}
ds^2 = a^2(\tau)\big(d\tau^2 - d\textbf{x}^2\big)
\end{equation}
(where $a(\tau)$ is the scale factor and $\tau$ is the conformal time), the equations of motion read

\begin{itemize} 
\item[]
\begin{equation}\hspace{-1.7cm}
    \ddot{\phi} + 2 \mathcal{H}\dot{\phi} -\nabla^2\phi+ a^2\frac{\partial V}{\partial \phi} =-\frac{1}{a^2}\bigg[\frac{g_{\phi A}}{f}\dot{\mathbf{A}}\cdot\nabla\times\mathbf{A} + \frac{g_{\phi X}}{f}\dot{\mathbf{X}}\cdot\nabla\times\mathbf{X}  + \frac{g_{\phi A X}}{f}\bigg(\dot{\mathbf{X}}\cdot\nabla\times\mathbf{A} + \dot{\mathbf{A}}\cdot\nabla\times\mathbf{X}\bigg)\bigg] \ ,
\end{equation} 

\item[]
\begin{equation}\hspace{-1.7cm}
\label{PhotonEq}
 \ddot{\mathbf{A}} + \sigma\bigg[\dot{\mathbf{A}} + \mathbf{v}\times(\nabla\times \mathbf{A})\bigg] + \nabla\times(\nabla\times \mathbf{A}) = \frac{g_{\phi A}}{f} \bigg(\dot{\phi} \nabla\times \mathbf{A} - \nabla\phi\times \dot{\mathbf{A}}\bigg) + \frac{g_{\phi A X}}{f} \bigg(\dot{\phi}\nabla\times \mathbf{X}  - \nabla\phi\times \dot{\mathbf{X}}\bigg) \ , 
\end{equation}
\item[]
\begin{equation}\hspace{-4.1cm}
     \ddot{\mathbf{X}} + \nabla\times(\nabla\times \mathbf{X})  = \frac{g_{\phi X}}{f} \bigg(\dot{\phi} \nabla\times \mathbf{X} - \nabla\phi\times \dot{\mathbf{X}}\bigg)+ \frac{g_{\phi A X}}{f} \bigg(\dot{\phi} \nabla\times \mathbf{A} - \nabla\phi\times \dot{\mathbf{A}}\bigg)  \ .
     \label{eqn:EoM}
\end{equation}
\end{itemize}

In the equations above, we imposed the temporal gauge $A_0 = 0$ and $X_0 = 0$, and overdots represent derivatives with respect to conformal time, while $\mathcal{H}$ is the conformal Hubble parameter (reading $\mathcal{H} = a H$, where $H$ is the Hubble parameter) and $\sigma$ is the conformal electric conductivity of the primordial plasma. Note that the term including $\sigma$ is proportional to the velocity of the plasma $\mathbf{v}$. A self-consistent treatment of such term requires to obtain $\mathbf{v}$ by solving the appropriate fluid equations, which must be coupled with the equations of motion for the axion-gauge field system. We defer such task to a future publication, and below we follow \citep{Choi_2018} and neglect such term. The conformal conductivity $\sigma$ is related to the physical conductivity $\sigma_{ph}$ via $\sigma_{ph} = \sigma/a$. We use the expression in \citep{Baym_1997} for $T_{QGP} \ll T \ll M_W$ (where $T_{QGP}$ is the temperature of the quark-gluon plasma and $M_W$ is the mass of the $W$ boson) given by

\begin{equation}
    \sigma_{ph} = \frac{J}{E} = \frac{N_l}{N_l + 3 \sum^{N_q}_q Q_q^2} \frac{3 \zeta(3)}{\ln (2) \alpha \ln(1/\alpha N_l)} T \ ,
    \label{eq:conduc_high}
\end{equation}
where $J$ is the electric current density, $E$ the electric field, $N_l$ and $N_q$ count, respectively, the lepton and quark species with masses below $T$, $Q_q$ is the quark charge in units of the electric charge $e$, and $\alpha = e^2/4\pi$. At lower temperatures, we use the expression \citep{Baym_1997}

\begin{equation}
    \sigma_{ph} = \frac{3 \zeta(3)}{\ln (2) \alpha \ln(1/\alpha)} T \ .
    \label{eq:conduc_low}
\end{equation} 

The temperature is evolved in time using the relation in \citep{Husdal_2016}, reading

\begin{equation}
T_{\textrm{MeV}} = \left(\frac{2.4}{\sqrt{g_*}t}\right)^{\frac{1}{2}} \ ,
\end{equation}
where $T_{\textrm{MeV}}$ is the universe temperature in MeV, $g_*$ denotes the energy density effective degrees of freedom, and $t$ is the cosmic time.

The system of coupled equations of motion leads to a complex, non-linear dynamics that requires lattice simulations. However, it is possible to get some insight by inspecting the source terms in the equations of motion. If $f$ is larger than the inflationary Hubble parameter $H_I$ and $T_R < f$, the initial axion field is homogeneous (pre-inflationary axions). If the axion is far from the minimum of its potential (misalignement mechanism), its energy density is several orders of magnitude larger than the initial electromagnetic energy density of the $\mathbf{A}$ and $\mathbf{X}$ fields, and its oscillations are damped by the expansion rate of the universe (given by the term $2\mathcal{H}\dot{\phi}$). When the Hubble parameter becomes smaller than the axion mass, the axion begins to oscillate. The time derivative of the axion field acts as a source term for the gauge fields (see the terms on the right-hand side of the corresponding equations of motion), producing the gauge fields via parametric resonance \cite{Kofman_1997, Amin_2015, Figueroa_2017, Agrawal_2018, Kitajima_2018, Choi_2018}, and leading to a \enquote{dynamo} effect that generates electric and magnetic fields\footnote{We note that if the axion is initially close to the minimum of its potential, its oscillations can induce perturbative particle production, cf. for example \citep{Ai_2023} and references therein.}. While the energy density of the dark-photon field grows resonantly, the plasma conductivity curbs the growth of the photon field due to axion oscillations, acting as a damping term similarly to the effect due to the expansion rate of the universe for the axion. For example, the term proportional to $g_{\phi A} \dot{\phi} \nabla\times \mathbf{A}$ does not lead to an efficient dynamo mechanism, since it depends on the weak, standard magnetic field $\nabla\times \mathbf{A}$. However, as shown below, it is possible to partially overcome the plasma-damping effect via the axion-photon-dark photon coupling terms, such as $g_{\phi A X} \dot{\phi} \nabla\times \mathbf{X}$. The latter term depends on the magnetic field associated with $\mathbf{X}$, rather than the standard, suppressed magnetic field, and allows for a more efficient production of standard electromagnetic fields. In general we find that while the dark electromagnetic fields are large, the standard electromagnetic fields remain subdominant due to the high plasma conductivity.

\subsection{Inhomogeneous axion field}
\label{sec:Fluctuations}
Inhomogeneities develop naturally for post-inflationary axions, i.e. when $f \lesssim H_I$. When the temperature of the universe is of the order of $T \sim f$, topological defects such as domain walls and networks of strings form via the Kibble mechanism \citep{Kibble_1976, Yamaguchi_1998, Yamaguchi_1998b, Yamaguchi_1999, Hiramatsu_2013, Kawasaki_2018, Gorghetto_2018, Buschmann_2020, OHare_2022, Pierobon_2023}, which release relativistic axions. Numerically, there are two possible approaches to study the evolution of the axion field in this scenario \citep{Pierobon_2023}. The first consists in solving the equation of motion for the Peccei-Quinn field, simulating the formation of topological defects and the production of relativistic axions, which later scale as non-relativistic matter. The second approach relies on the construction of initial conditions for axions from the string network spectrum, and the simulations start without the presence of topological defects, but with an axion field whose degree of inhomogeneity depends on the properties of the topological defects. 

However, even for pre-inflationary axions there can be conditions in which the axion field becomes inhomogeneous. For example, if the reheating temperature $T_R$ (with $T_R\lesssim \sqrt{H_I M_{Pl}}$, where $M_{Pl}$ is the Planck mass) after the inflationary period is larger than the axion decay constant, the global symmetry associated with the axion is restored after inflation. At later times, the breaking of such symmetry leads to the formation of a cosmic network of strings, and the resulting axion field is inhomogeneous. 

To study how initial inhomogeneities affect the ability of the axion to source gauge fields, we perform simulations of axions with decay constants in the range $f \in [10^{11}, 10^{15}]$ GeV. Such a range encompasses scenarios in which inhomogeneities of the initial axion field develop when $f \lesssim H_I$, or when $T_R \gtrsim f$. We use simplified initial conditions, which are meant to reproduce (at least qualitatively) the complex axion field configurations due to a network of strings and domain walls \citep{Yamaguchi_1998, Yamaguchi_1998b, Yamaguchi_1999, Vaquero_2019, Buschmann_2020, OHare_2022, Pierobon_2023}. We initialize the axion field drawing a random value of the field for each Hubble patch from a gaussian distribution \citep{Enander_2017}. We assume that the field is homogeneous on sub-Hubble scales, which corresponds to considering only the misalignement contribution for the axion field \citep{Enander_2017, Vaquero_2019}. 
We stress that such initial conditions are employed in order to study qualitatively how the gauge field production mechanism responds to an initial, highly inhomogeneous field configuration. Moreover, the typical resolution employed in the literature for simulations with post-inflationary axions (without gauge fields) is $> 1000^3$ grid points, which are necessary to resolve properly the formation of small-scale axion structures. Running such simulations demands a significant computational power, which increases further if the axion interacts with multiple gauge fields, which is our primary goal. To the best of our knowledge, there are no numerical campaigns in the literature where such high-resolution simulations are employed to study inhomogeneous axions interacting with two gauge fields. The capabilities of the numerical code that we developed do not allow us to perform runs in a reasonable time including axions and gauge fields with such resolutions, and rather allow us to use the standard resolution employed in the literature of axions interacting with dark photons only, typically employing $128^3$ and $256^3$ grid points. While the limited resolution of our runs makes our results for initially inhomogeneous axion fields qualitative, we aim at performing higher-resolution simulations which will be presented in future work.

\section{Lattice simulations}
\label{sec:lattice_sim}

We now report the details of the lattice simulations performed in this work. We consider the ALP in our model to be slightly lighter than the QCD axion at zero temperature, i.e. $m_\phi = 0.8 m_{QCD}(f)$.  In all of the following simulations, the gauge fields are initialized in momentum space as in \citep{Agrawal_2018} (albeit dark photons are massless in our model), with root-mean square amplitude in Fourier space

\begin{equation}
\sqrt{\langle|A_\mathbf{k}|^2\rangle} = \frac{1}{\sqrt{2 |k|/a}} \ ,
\end{equation}
where $k$ denotes the comoving momentum. The same initialization procedure is applied to the dark photon field.
The simulations are performed in a cubic volume with linear dimension $L = \pi/(4 m_\phi)$ using $128^3$ grid points, applying periodic boundary conditions at the edges of the simulation box and starting at the conformal time $\tau_i = 0.1L$. We follow \citep{Agrawal_2018} and evolve the system of equations in time using a leap-frog algorithm and an implicit scheme where the terms multiplying the updated values of the field time-derivatives are inserted in a matrix $\mathcal{M}^{l m}$. We solve the system $\mathcal{M}^{l m}V_m(\tau+\Delta \tau) = Z_l$, where $V_m$ is a vector containing the updated values of the time-derivatives of the fields, $Z_l$ is a vector containing the values of the time-derivatives at the previous time step, and $\Delta\tau$ is the conformal time-step. We also check that the Gauss constraint \citep{Kitajima_2018, Ratzinger_2021} is satisfied in all of the runs.

\subsection{Magnetogenesis when $T_R < f$}
\label{sec:pre_infl}

\begin{figure*}[t!]
\hspace{-15mm}
\includegraphics[width=17.3cm, height = 6.3cm]{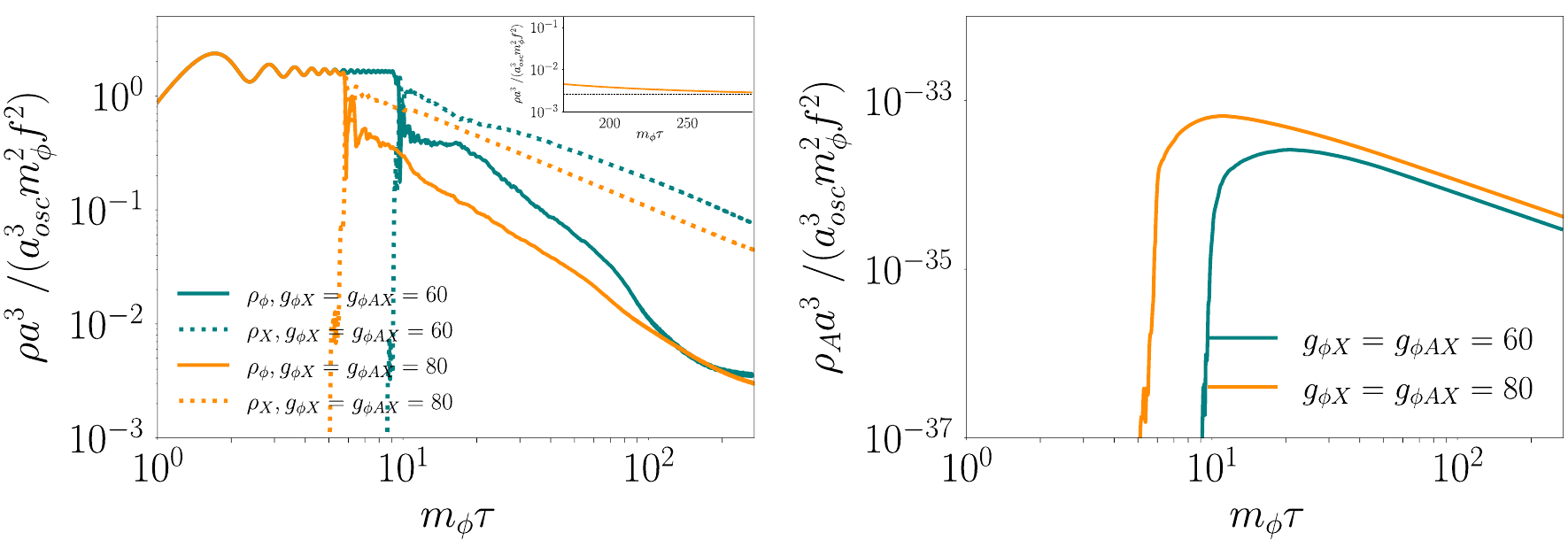}%
\caption{Volume-averaged energy densities for axions (with $f = 10^{14}$ GeV), dark photons and photons ($\rho_\phi, \rho_X$ and $\rho_A$ respectively) versus conformal time. Left panel: axion and dark photon energy densities for $g_{\phi X} = g_{\phi A X} = 60$ and $g_{\phi X} = g_{\phi A X} = 80$. The inset, smaller panel displays (in linear scale for $m_{\phi}\tau$) that the orange solid line scales as $a^{-3}$ at later times with respect to the teal curve due to the stronger coupling. Right panel: photon energy density for the same couplings.
In all of the panels, the axion-photon coupling is fixed to $g_{\phi A} = -2.2\times 10^{-3}$, and the axion mass is $m_\phi = 4.7 \times 10^{-8}$ eV. The initial value of the axion field is $\phi = f$. }
\label{fig_energy_dens_ALP}
\end{figure*}

We show results of our simulations with $f = 10^{14}$ GeV in Figure \ref{fig_energy_dens_ALP}, assuming that $T_R < f$ and hence that the axion field is initially homogeneous with initial amplitude $\phi = f$.
In the left panel of Figure \ref{fig_energy_dens_ALP} we show the volume-averaged energy densities of ALPs and dark photons (denoted with $\rho_\phi$ and $\rho_X$ respectively) normalized by $m^2_\phi f^2$ and multiplied by $(a/a_{osc})^{3}$, where we define $a_{osc}$ as the value of the scale factor when $H = m_\phi$. The axion-photon coupling is fixed throughout this paper to $g_{\phi A} = -2.2\times 10^{-3}$ (as for the KSVZ axion model \citep{Cortona_2016}), while we vary the remaining couplings and use $g_{\phi X} = g_{\phi A X} = 60$ (teal curves) and $g_{\phi X} = g_{\phi A X} = 80$ (orange curves). At the beginning of the simulation, the axion is pinned by the Hubble term. Later, the axion starts to oscillate and its energy density $\rho_\phi$ scales as $a^{-3}$ until $\rho_X$ becomes comparable to $\rho_\phi$. Then, the latter drops sharply due to the backreaction of the sourced dark photon field, and the axion field develops significant inhomogeneities. After a transient period in which axions and dark photons keep interacting, at the end of the simulations the axion energy density decreases as $a^{-3}$ (i.e. axions behave as cold dark matter), while the energy density of dark photons scales as $a^{-4}$ (i.e. the gauge fields behave as dark radiation). The stronger is the coupling, the shorter is the conformal time interval required for the energy density of dark photons to become higher than the one of the axion. The right panel in Figure \ref{fig_energy_dens_ALP} displays the evolution of the volume-averaged, standard electromagnetic energy density $\rho_A$ for the coupling values employed in the other panel. We find that $\rho_A$ is far smaller than $\rho_\phi$ and $\rho_X$, attaining a peak of $\rho_A (a/a_{osc})^{3}/m^2_\phi f^2\approx 6 \times 10^{-34}$ for $g_{\phi X} = g_{\phi A X} = 80$. While the growth of the dark photon field takes place via the conversion of axion energy density into dark photons (with efficiency regulated by the magnitude of the coupling $g_{\phi X}$), the ALP field does not source photons efficiently. As described in the previous section, the photon oscillations are damped by the highly conductive nature of the primordial plasma, the standard magnetic field $\nabla \times \mathbf{A}$ remains small, and the $g_{\phi A} \dot{\phi} \nabla\times \mathbf{A}$ source term is inefficient even for $g_{\phi A} \approx \mathcal{O}(100)$. However, when the dark photon abundance increases due to the axion-driven dynamo, the coupling $g_{\phi A X}$ allows for a secondary dynamo effect that sources standard electric and magnetic fields. This secondary dynamo is more efficient than the one due to the axion-photon coupling because the corresponding source term in the photon equation of motion (Eq. \eqref{PhotonEq}) is proportional to $g_{\phi A X} \dot{\phi} \nabla\times \mathbf{X}$, with $|\nabla\times \mathbf{X}| \gg |\nabla\times \mathbf{A}|$. After the exponential growth of $\rho_A$, the latter scales as $a^{-4}$. We note that the coupling $g_{\phi A X} $ in our model is larger than the one employed in \cite{Choi_2018}. This is because the constraints on the product of the axion-photon-dark photon constant and the sourced magnetic for ultra-light axions (with masses $\lesssim 10^{-14}$ eV) are stricter than for heavier axions \citep{Tashiro_2013}.

\begin{figure*}
\hspace{-5mm}
\includegraphics[width=7.8cm, height = 6.8cm]{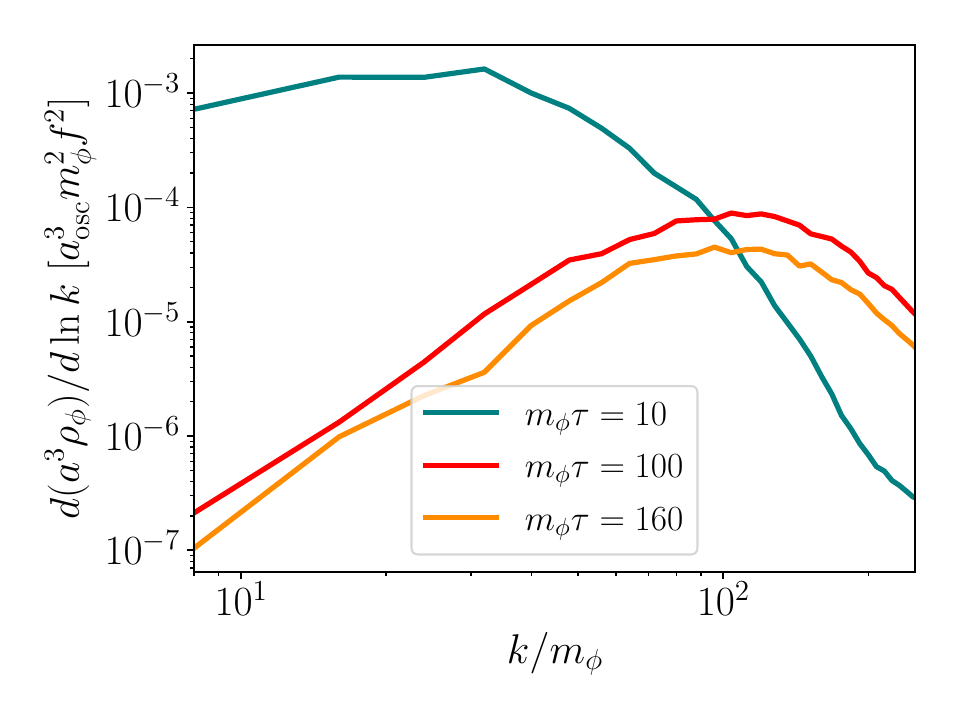}%
\includegraphics[width=7.8cm, height = 6.8cm]{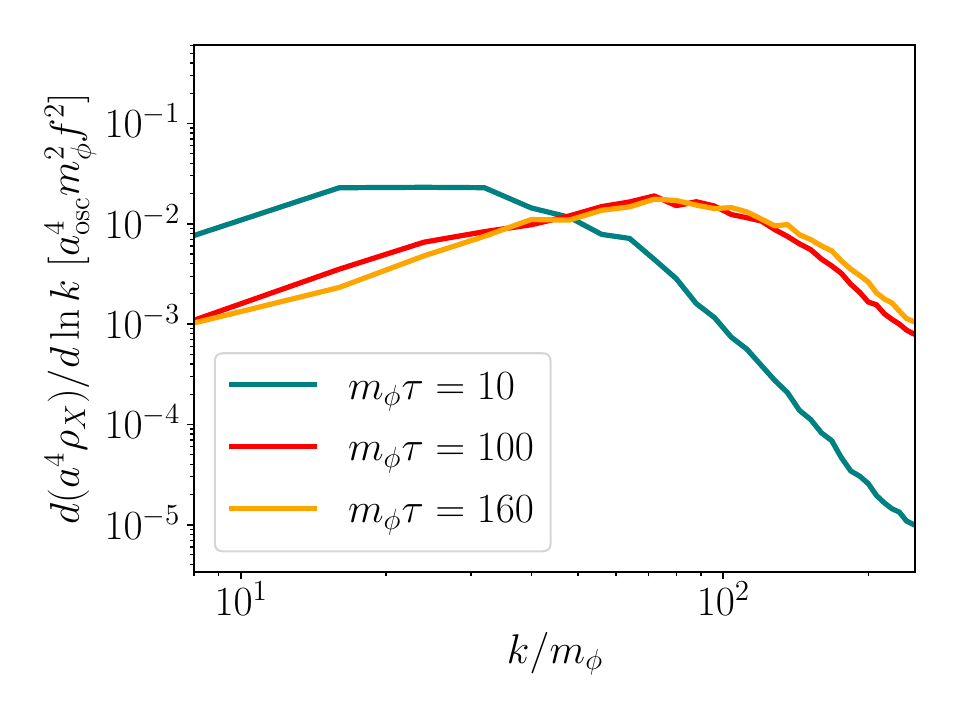}
\caption{Spectra of axions (left panel) and dark photons (right panel) for $f = 10^{14}$ GeV, $m_{\phi} = 4.7 \times 10^{-8}$ eV and $g_{\phi X} = g_{\phi A X} = 60$.}
\label{fig_Spectra_ALP}
\end{figure*}

\begin{figure*}
\hspace{18mm}
\includegraphics[width=9.5cm, height = 6.5cm]{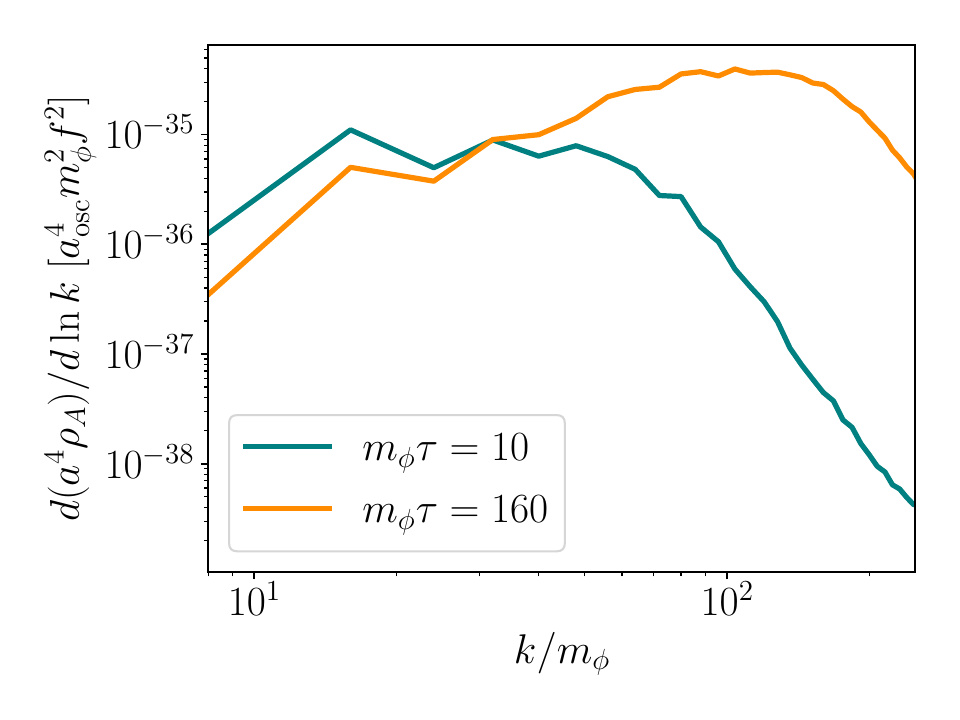}%
\caption{Spectrum of the photon energy density. The axion parameters and couplings are as in Figure \ref{fig_Spectra_ALP}.}
\label{fig_Spectrum_photon_f14_g60}
\end{figure*}

\begin{figure*}
\hspace{-5mm}
\includegraphics[width=7.8cm, height = 7cm]{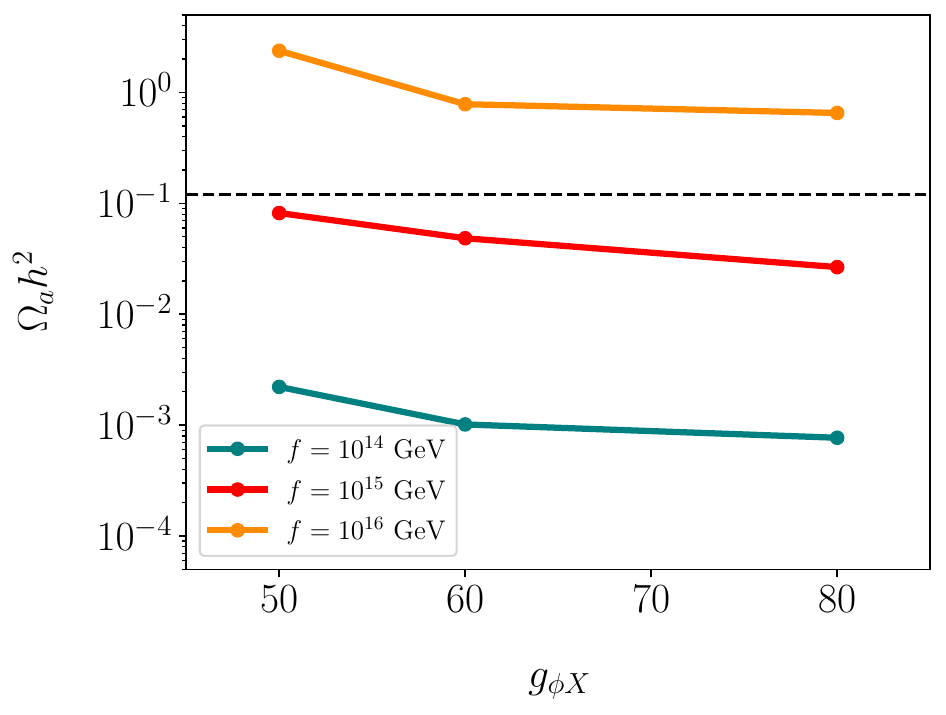}%
\includegraphics[width=8cm, height = 7.2cm]{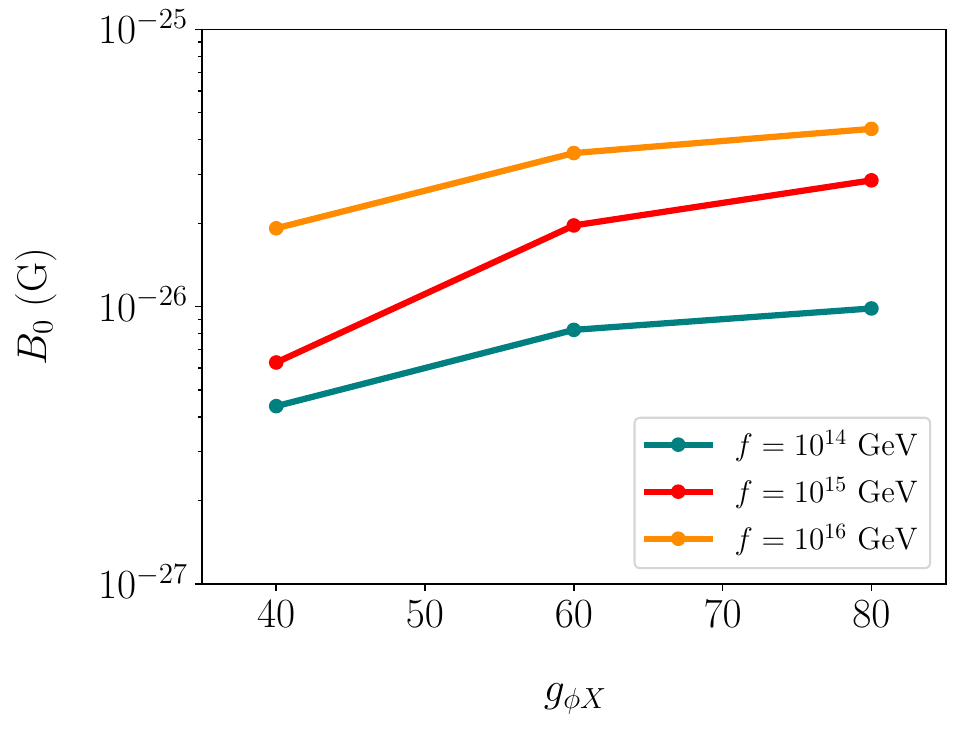}
\caption{Relic axion energy density versus the coupling $g_{\phi X}$ (left panel) and standard magnetic field strength redshifted to today (right panel). The mass of the axion is fixed to $m_\phi = 4.7 \times 10^{-8}$ eV (for $f = 10^{14}$ GeV), $m_\phi = 4.7 \times 10^{-9}$ eV (for $f = 10^{15}$ GeV) and $m_\phi = 4.7 \times 10^{-10}$ eV (for $f = 10^{16}$ GeV). The results are obtained with $\phi = f$ at the start of the simulations, and with $g_{\phi X} = g_{\phi A X}$.}
\label{fig_relics_pre}
\end{figure*}

In Figure \ref{fig_Spectra_ALP} we show the axion spectra (left panel), as well as the dark photon spectra (right panel) for the runs in Figure \ref{fig_energy_dens_ALP}. In both panels, the spectra show a single peak which is first located at $k/m_{\phi} \approx 30$ and later moves to higher momenta ($k/m_{\phi} \gtrsim 100$) \citep{Ratzinger_2021}. As the conformal time increases, one can see how the power transfers to higher momenta due to the scattering of low-momentum axions with high-momentum dark photons, leading to the production of axions and dark photons with higher momenta and hence to the growth of the spectra for $k/m_{\phi} \gtrsim 80$. The spectrum of the standard electromagnetic energy density is reported in Figure \ref{fig_Spectrum_photon_f14_g60}. The spectral features are similar to the ones of the dark photon spectra, and display a transfer of power from smaller to larger momenta at late times.

In Figure \ref{fig_relics_pre} we show the axion relic density $\Omega_ah^2$ (left panel) and the strength of the standard magnetic field redshifted to today $B_0$ (right panel) versus the axion-dark photon coupling (we use $g_{\phi X} = g_{\phi A X}$). In the left panel, the axion relic density $\Omega_ah^2$ approaches the value $\Omega_{DM} h^2 = 0.12$ (horizontal black line) for $f = 10^{15}$ GeV, while it is lower or higher for $f = 10^{14}$ GeV and $f = 10^{16}$ GeV respectively. The right panel shows that, as expected, the higher is the coupling, the more intense is the relic magnetic field strength. Note that while $B_0$ is maximal for $f = 10^{16}$ GeV, it corresponds to having $\Omega_a h^2 \gg 0.12$; hence, only for $f \leq 10^{15}$ GeV we obtain that $\Omega_a h^2$ is compatible with $\Omega_{DM} h^2$, and at the same time we find a sizable $B_0$. We stress however that due to the typical axion masses studied in this section, the correlation length of the magnetic fields in our simulations (of the order of $1$-$10$ pc) is far shorter than the typical correlation lengths of large-scale magnetic fields. In the next section however we consider lighter axions and hence larger correlation lengths for the sourced magnetic fields.

\subsection{Magnetogenesis when $T_R > f$}
\label{sec:fluct_lattice}
We now study the production of gauge fields in the presence of an initially inhomogeneous axion field, fixing $f = 10^{14}$ GeV and assuming $T_R \gtrsim f$, as outlined in Section \ref{sec:Fluctuations}. We introduce a parameter $\zeta$ that quantifies the ratio between the \textit{initial} total energy density of the axion (i.e. gradient plus potential energy densities) and the magnitude of the potential energy of the axion given by $m^2_\phi f^2$, i.e. $\zeta = \rho_\phi(\tau_i) / (m^2_\phi f^2)$. As $\zeta$ increases, so does the initial average value of the axion field, leading to a higher axion relic density\footnote{We note that simulations of post-inflationary axions report relic densities which can be far larger than the one obtained from the standard misalignement mechanism for pre-inflationary axions (cf. \citep{Marsh_2016} and references therein).}. For $\zeta \approx 10^6$, the average value of the axion field is $\langle \phi^2/f^2 \rangle \approx \pi^2/3$ \citep{Marsh_2016}. As in Section \ref{sec:pre_infl}, we start the simulations at $\tau_i = 0.1L$, which in the scenario considered in this section corresponds to the moment when relativistic axions are emitted from the topological defect network and start interacting with the gauge fields.

\begin{figure*}[h]
\hspace{-10mm}
\includegraphics[width=17.cm, height = 11.5cm]{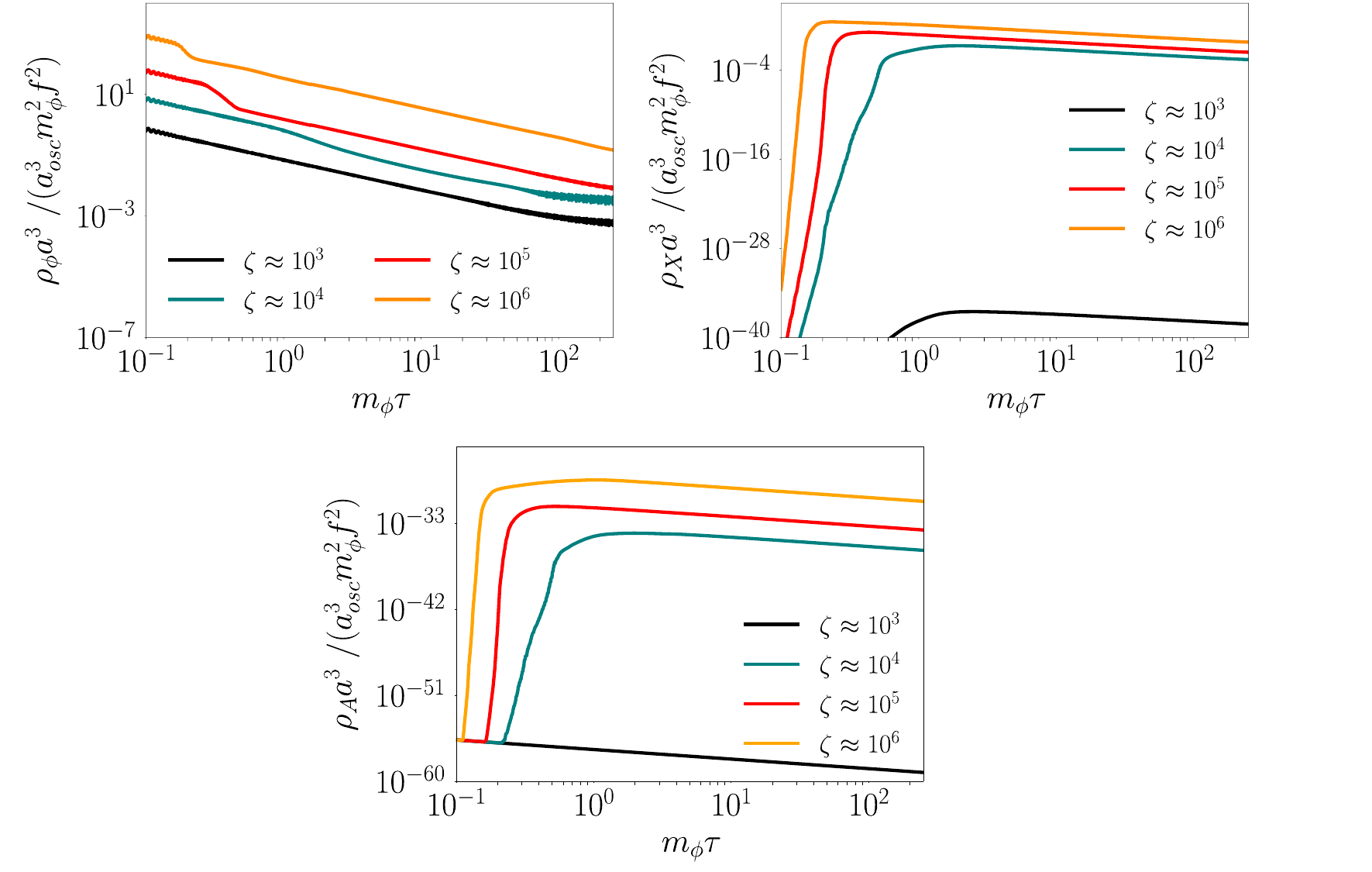}%
\caption{Production of gauge fields with $f = 10^{14}$ GeV, $m_{\phi} = 4.7 \times 10^{-8}$ eV, $g_{\phi X} = g_{\phi A X} = 60$ and assuming that $T_R > f$. The top-left panel shows that the axions scale as radiation until $m_\phi \tau \approx 100$, i.e. until the gradient energy is much larger than the potential term. The top-right panel shows that $\rho_{X} > \rho_{\phi}$ only for $\zeta \gtrsim 10^{4}$. Similarly, the bottom panel shows that there is no production of standard electromagnetic fields for $\zeta \lesssim 10^{3}$.}
\label{fig_energy_dens_String_f14}
\end{figure*}

Figure \ref{fig_energy_dens_String_f14} reports the energy densities of axions, dark photons and photons in the top left, top right and bottom panels respectively for three different values of $\zeta$ and for $g_{\phi X} = g_{\phi A X} = 60$. At the beginning of the simulations, the top left panel shows that $\rho_\phi$ scales as $a^{-4}$, i.e. axions behave as radiation. This is because the initial gradient energy of the axion is far larger than the potential term, and the axion behaves as a nearly massless field. While $\rho_\phi$ is initially far larger than in the scenario studied in Section \ref{sec:pre_infl} ($\zeta \gg 1$), it quickly approaches the values attained in Figure \ref{fig_energy_dens_ALP}. Once the gradient energy density of the axion becomes comparable or smaller than its potential energy density, $\rho_\phi$ scales as $a^{-3}$, behaving like dark matter. The higher is $\zeta$, the longer it takes for the axion field to scale as $a^{-3}$ (cf. the black and teal curves with the red and orange ones). The top right panel of Figure \ref{fig_energy_dens_String_f14} shows that the production of dark photons is far less efficient when the axion field is inhomogeneous and, for $\zeta \approx 10^3$, $\rho_X$ remains well below $\rho_\phi$. The high degree of inhomogeneity of the axion field prevents the latter from acting as a coherently oscillating condensate, limiting the energy transfer from the axion to the gauge sector compared to when $T_R < f$ for the same couplings (cf. Figure \ref{fig_energy_dens_ALP}). The situation changes when $\zeta \gtrsim 10^{4}$, allowing $\rho_X$ to become larger than $\rho_\phi$. The bottom panel shows the growth of $\rho_A$. The latter becomes a few orders of magnitude larger than in Figure \ref{fig_energy_dens_ALP} when $\zeta \gtrsim 10^4$, signalling that an inhomogeneous axion field has a beneficial effect on the magnitude of the produced standard magnetic field. Thanks to the large axion gradients, the source terms in the gauge field equations of motion lead to a more efficient magnetogenesis despite the suppression due to the plasma conductivity. 

\begin{figure*}[h]
\hspace{-11mm}
\includegraphics[width=16.5cm, height = 6.5cm]{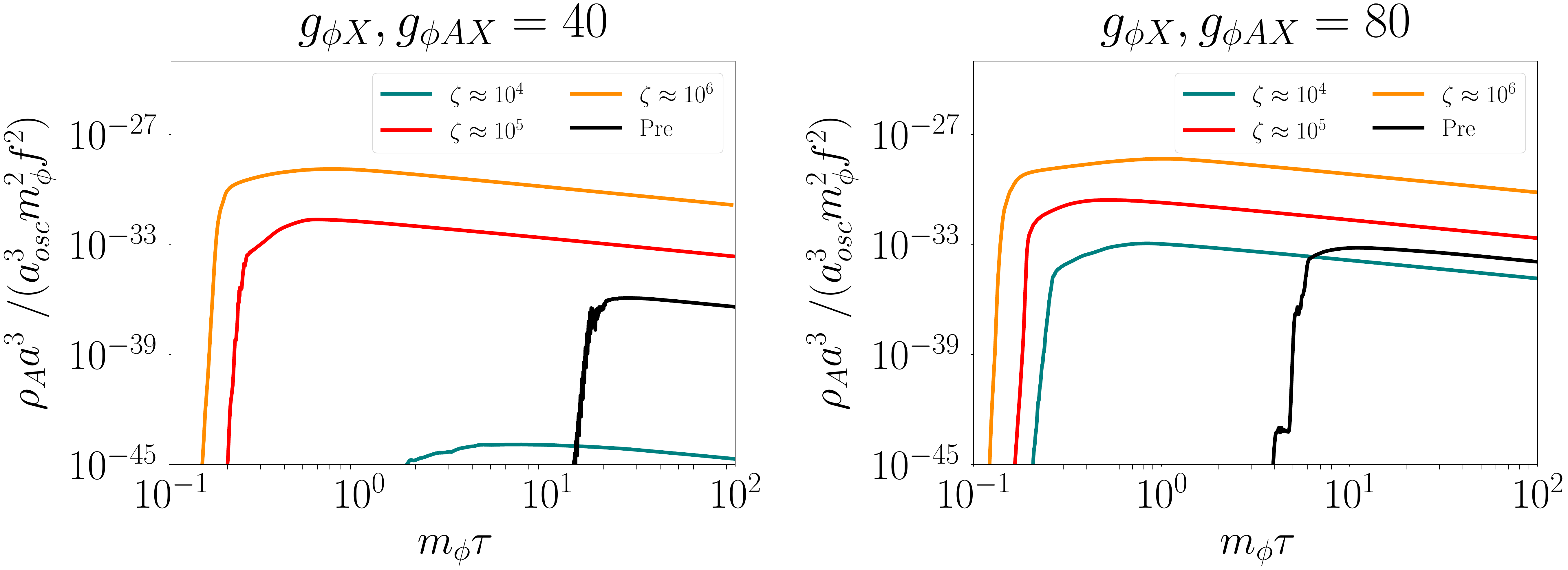}%
\caption{Comparison of the standard electromagnetic energy produced by pre-inflationary axions (black curves, obtained with the initial condition $\phi = f$) and an initially inhomogeneous axion field for $f = 10^{14}$ GeV and $m_{\phi} = 4.7 \times 10^{-8}$ eV. The left panel is obtained for $g_{\phi X} = g_{\phi A X} = 40$, while the right panel for $g_{\phi X} = g_{\phi A X} = 80$.}
\label{fig_Pre_post_magnetogenesis}
\end{figure*}

\begin{figure*}
\hspace{19mm}
\includegraphics[width=9.5cm, height = 6.5cm]{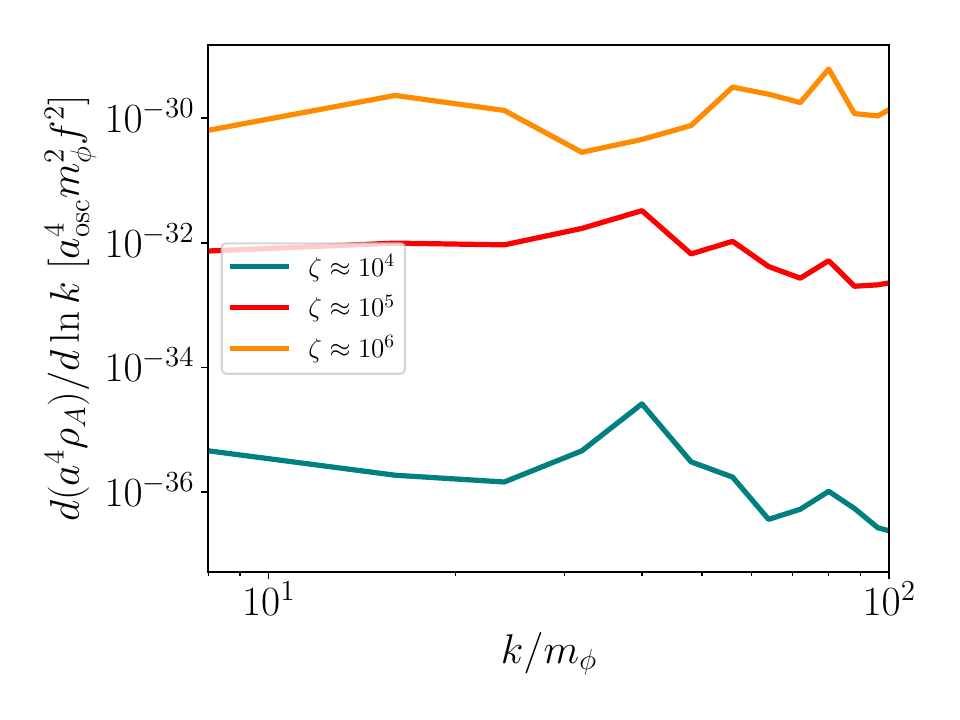}%
\caption{Spectrum of the standard electromagnetic energy density for an initially inhomogeneous axion field with parameters fixed to $f = 10^{14}$ GeV , $m_{\phi} = 4.7 \times 10^{-8}$ eV and with $g_{\phi X} = g_{\phi A X} = 80$.}
\label{fig_Spectrum_string_photon}
\end{figure*}

In Figure \ref{fig_Pre_post_magnetogenesis} we compare $\rho_A$ obtained for pre-inflationary axions (black curves, with the initial condition of the homogeneous axion field given by $\phi = f$) with results obtained for $T_R > f$. For both the $g_{\phi X} = g_{\phi A X} = 40$ and $g_{\phi X} = g_{\phi A X} = 80$ cases, when $\zeta \gtrsim 10^5$ the production of standard electromagnetic fields is boosted, exceeding by a few orders of magnitude the results obtained with pre-inflationary axions. In particular, we find that $\rho_A$ obtained for $\zeta \approx 10^{6}$ is four orders of magnitude larger than the values attained by the black curves. We report additional results for initially inhomogeneous axions in Appendix \ref{sec:appendix_A}. We also confirm the magnetogenesis boost due to inhomogeneous axions by performing a set of short simulations using $256^3$ and $320^3$ grid points and $f = 10^{14}$ GeV, which are reported in Appendix \ref{sec:higher_res} along with the spectra of $\rho_X$. The spectra of $\rho_A$ obtained from such runs are reported in Figure \ref{fig_Spectrum_string_photon}, where we fix the couplings to $g_{\phi X} = g_{\phi A X} = 80$ and vary $\zeta$. The teal and red curves peak at around $k/m_{\phi} \approx 40$, while the orange curve shows a peak at $k/m_{\phi} \approx 80$. The latter curve also shows a bump at relatively low momenta ($k/m_{\phi} \approx 20$) compared to the flatter slope of the spectra for $\zeta \approx 10^4$ and $\zeta \approx 10^5$. We emphasize that the results in this section should be interpreted qualitatively, and higher-resolution runs are required to resolve properly the growth of $\rho_X$, $\rho_A$ and their spectral features.

\begin{figure*}
\hspace{-5mm}
\includegraphics[width=7.8cm, height = 6.8cm]{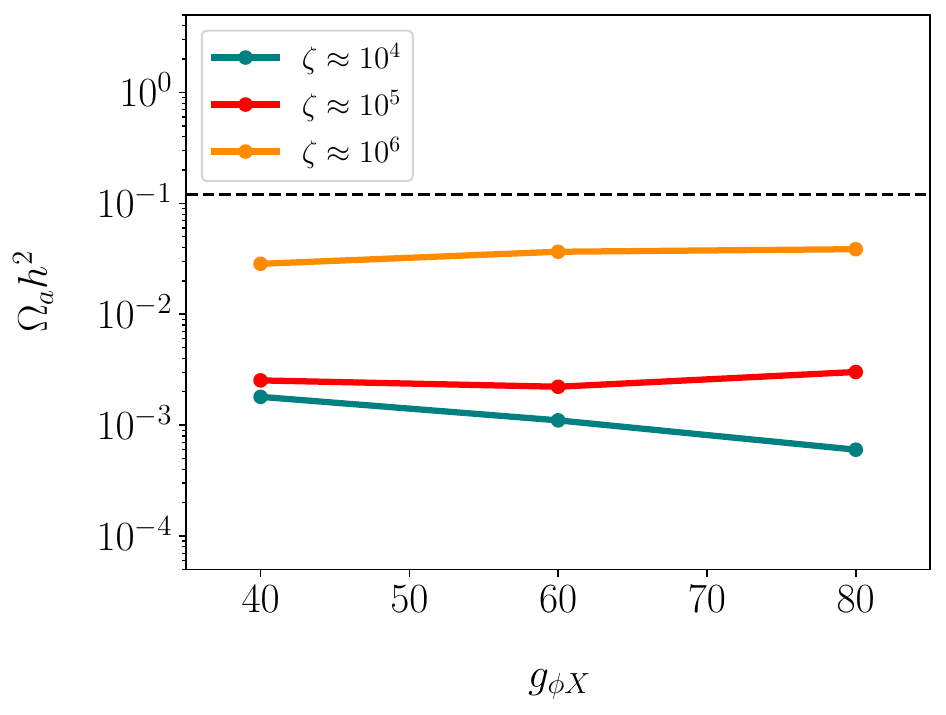}%
\includegraphics[width=7.8cm, height = 6.8cm]{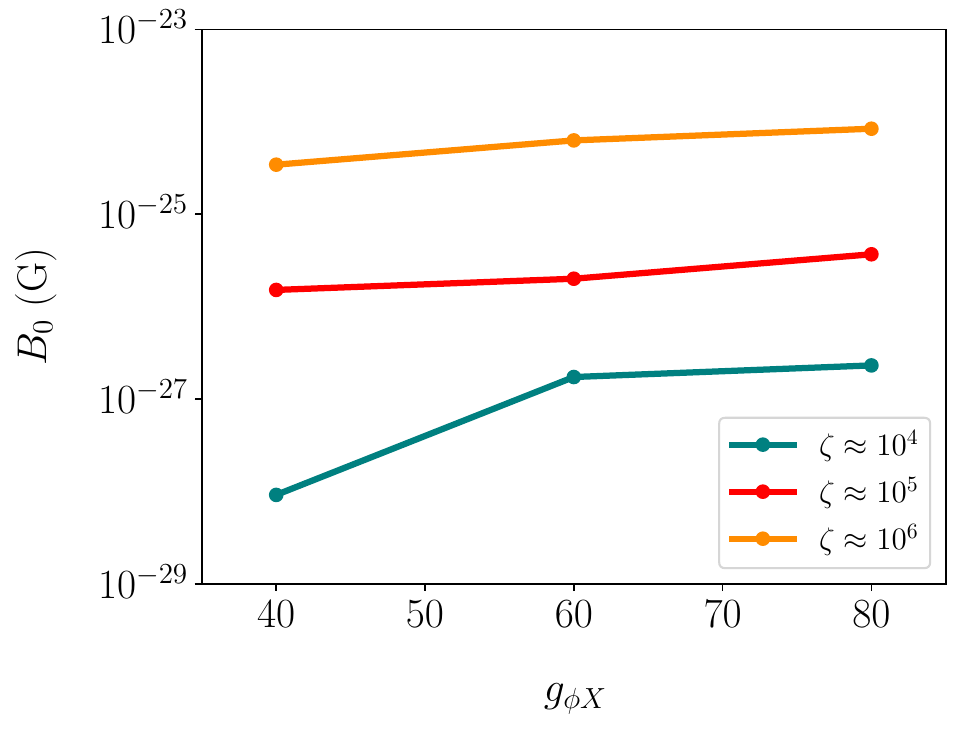}
\caption{Relic axion density versus the coupling $g_{\phi X}$ (left panel) and standard magnetic field strength redshifted to today (right panel) for axions with $f = 10^{14}$ GeV, $m_{\phi} = 4.7 \times 10^{-8}$ eV and assuming that $T_R > f$. We also use $g_{\phi X} = g_{\phi A X}$.}
\label{fig_relics_post}
\end{figure*}

Figure \ref{fig_relics_post} reports the relic axion density (left panel) and the standard magnetic field strength redshifted to today (right panel) versus $g_{\phi X}$ (we use $g_{\phi X} = g_{\phi A X}$). We find that for $\zeta \lesssim 10^5$, $\Omega_a h^2$ is well below $\Omega_{DM} h^2$ (given by the horizontal black line), and attains values similar to the ones found in the left panel of Figure \ref{fig_relics_pre} for $f = 10^{14}$ GeV. On the other hand, for $\zeta \approx 10^6$ we find that $\Omega_a h^2$ approaches $\Omega_{DM} h^2$. The right panel shows that, for $\zeta \approx 10^6$, $B_0$ is two orders of magnitude larger than in the scenario studied in Figure \ref{fig_relics_pre} for $f = 10^{14}$ GeV, with a maximum strength of $B_0 \approx 8 \times 10^{-25}$ G, which is close to the value obtained for the case in which magnetogenesis takes place after the Big Bang Nucleosynthesis \citep{Choi_2018}.

In general, the results in this section show that if the axion is initially inhomogeneous ($T_R > f$), the production of gauge fields is severely hampered for $\zeta \lesssim 10^4$ and $g_{\phi X}, g_{\phi A X} \lesssim 60$. In this case dark photons remain far less abundant than the axion, in contrast to the results in Section \ref{sec:pre_infl}, which correspond to the scenario thoroughly studied in the literature \citep{Agrawal_2018, Kitajima_2018, Ratzinger_2021}. However, inhomogeneous axions can lead to stronger magnetic fields, which in the case of photons are roughly two orders of magnitude more intense than in the scenario $T_R < f$, at least for high values of $\zeta$ and relatively large couplings. 

\begin{figure*}
\hspace{20mm}
\includegraphics[width=9.5cm, height = 7.cm]{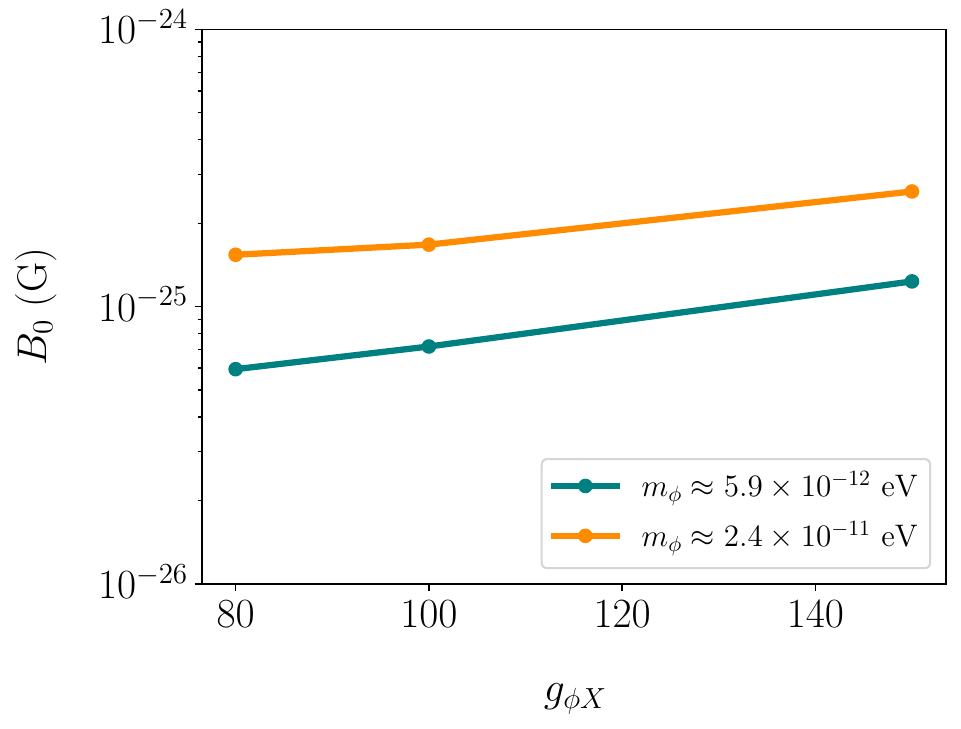}%
\caption{Standard relic magnetic field strength for ultra-light axions with $f = 10^{15}$ GeV and for $m_{\phi} = 5.9 \times 10^{-12}$ eV (teal curve) and $m_{\phi} = 2.4 \times 10^{-11}$ eV (orange curve). The relic magnetic field strength is obtained using $g_{\phi X} = g_{\phi A X}$ and $\zeta \approx 10^6$.}
\label{fig_relic_mag_field_post_ULA}
\end{figure*}

In the results above the typical magnetic field correlation length is smaller than the one characteristic of large-scale, cosmological magnetic fields due to our choice of axion parameters. It is possible to obtain larger correlation lengths by considering lighter axions. Figure \ref{fig_relic_mag_field_post_ULA} shows the standard relic magnetic field strength for axions with $f = 10^{15}$ GeV, $m_\phi \approx 5.9 \times 10^{-12}$ eV\footnote{Note that the axion parameters $m_\phi \approx 5.9 \times 10^{-12}$ eV and $f = 10^{15}$ GeV are in principle excluded by the recent bounds on $f$ and $m_\phi$ obtained in \citep{Mehta_2020}, where the constraints are obtained from compactifications of IIB string theory (cf. also \citep{AxionLimits} for current bounds on $m_\phi$ and $f$). Here we consider such a combination of $m_\phi$ and $f$ for illustration purposes only.} and $m_\phi \approx 2.4 \times 10^{-11}$ eV, and $\zeta \approx 10^6$ (assuming $T_R > f$ and hence that the axion field is initially inhomogeneous). With such parameters, the magnetic field has a correlation length of order $\mathcal{O}(0.1)$ kpc, and attains $B_0 \approx 10^{-25}$ G for the relatively large couplings $g_{\phi X} = g_{\phi A X} = 150$. Note that the sourced magnetic field is weaker with respect to the intensities reported in the right panel of Figure \ref{fig_relics_post}. This is mostly due to the numerical initialization procedure employed in our code (cf. Appendix \ref{sec:appendix_A}).

%%\appendix
%%\section{Some title}
%%Please always give a title also for appendices.

\section{Conclusion}

The parameter space of axions is restricted by the $\Lambda$CDM model, which gives a relic dark matter density of $\Omega_{DM}h^2 = 0.12$ \cite{Planck_2020}. This translates into bounds on the initial axion field amplitude, mass and on the axion decay constant.

When the axion field is initially homogeneous, we find that the production of dark photons and photons via axion oscillations alleviates the tension between the relic density of axions with large decay constants and $\Omega_{DM}$, and allows for the production of small-scale, standard magnetic fields with strengths redshifted to today of the order of $10^{-27}$-$10^{-26}$ G and correlation lengths of the order of $1$-$10$ pc. We also perform the first simulations where the initial axion field is highly inhomogeneous, which is typical for post-inflationary axions (i.e. axions with $f < H_I$), while for axions with large $f$ these conditions are possible if $T_R > f$. In such a scenario, the axion relic density is close to $\Omega_{DM}h^2$ for $f = 10^{14}$ GeV and $\zeta \approx 10^6$. We also show that a cosmologically relevant abundance of dark photons is more challenging to achieve, unless the $g_{\phi X}$ coupling is large (with $g_{\phi X} \gtrsim 60$) and/or $\zeta \gtrsim 10^4$. On the other hand, an initially inhomogeneous axion field has the beneficial effect of boosting the strength of the sourced standard magnetic field by two orders of magnitude, if the axion gradients and the axion-photon-dark photon coupling are relatively large. In this case, the strength of the standard magnetic field redshifted to today attains $\approx 10^{-25}$ G, both in the case of ALPs with masses close to the QCD axion, and for ultra-light ALPs with masses of the order of $10^{-12} - 10^{-11}$ eV. For the latter, the typical correlation length is of the order of $\approx 0.1$ kpc, showing that it is possible to obtain large-scale magnetic field seeds that are relevant for galactic scales.

The present work can be expanded in several ways. Among them, it is desirable to improve the initial conditions for inhomogeneous axions employed in this work, for example along the lines of \citep{Pierobon_2023}. Moreover, we aim at performing higher-resolution runs with $512^3$ grid points or more to resolve properly the axion-gauge field dynamics and their spectra when the axion field is initially inhomogeneous.

\acknowledgments
It is a pleasure to thank Liina Jukko for providing us with the code to compute the spectra and David J. E. Marsh, Wen-Yuan Ai and Andrew Melatos for many useful comments and discussions. FA thanks Jun'ichi Yokoyama, Ryusuke Jinno and the RESCEU group at the University of Tokyo, and Shuichiro Yokoyama, Kiyotomo Ichiki and the Kobayashi-Maskawa Institute for hospitality while part of this research was completed. FA is supported by the Australian Research Council (ARC) Centre of Excellence for Gravitational Wave Discovery (OzGrav), through project number CE170100004. The numerical simulations presented in this work were performed on the Ngarrgu Tindebeek supercomputing cluster at Swinburne University.

\begin{figure*}[h]
\hspace{-20mm}
\includegraphics[width=19.5cm, height = 14.cm]{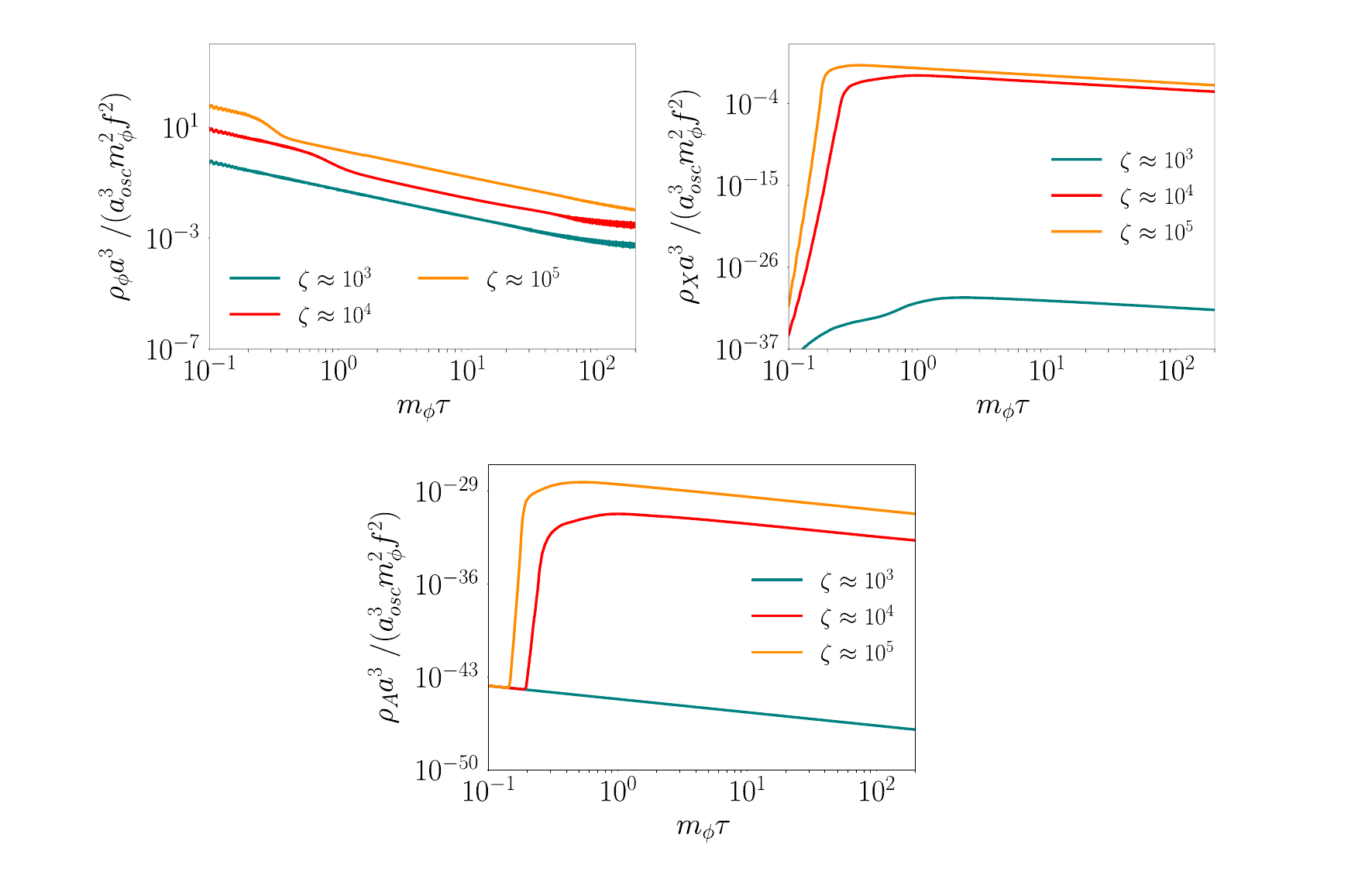}%
\caption{Production of gauge fields by post-inflationary axions with $f = 10^{11}$ GeV, $m_{\phi} = 4.7 \times 10^{-5}$ eV and using $g_{\phi X} = g_{\phi A X} = 60$. Axions scale as radiation until $m_\phi \tau \approx 100$. The top-right panel shows that the production of dark photons is efficient only for $\zeta \gtrsim 10^{4}$, and the bottom panel shows that standard electromagnetic fields are not produced for $\zeta \lesssim 10^{3}$.}
\label{fig_energy_dens_String_f11}
\end{figure*}

\appendix
\section*{Appendix}

\section{Additional results for inhomogenous axions}
\label{sec:appendix_A}

In Section \ref{sec:fluct_lattice} we report results for an initially inhomogeneous axion field using $f = 10^{14}$ GeV, which requires to have $T_R > f$. Here we study the case of axions with $f = 10^{11}$ GeV, which can be labeled as post-inflationary axions if $f < H_I$. Figure \ref{fig_energy_dens_String_f11} displays $\rho_\phi$, $\rho_X$ and $\rho_A$, which show a similar behaviour to Figure \ref{fig_energy_dens_String_f14} in Section \ref{sec:fluct_lattice}. The main difference can be seen in the bottom panel: $\rho_A$ attains a larger maximum value than in Figure \ref{fig_energy_dens_String_f14}, which is due to the different initial conditions for the gauge fields employed for runs with $f = 10^{11}$ GeV and $f = 10^{14}$ GeV. Numerically, our initialization procedure follows the one in the publicly available code LATTICEEASY \citep{Felder_2008}, so the initial magnitude of the gauge fields depends on the axion parameters (besides the size of the simulation box and the grid-spacing).

\section{Higher resolution runs}
\label{sec:higher_res}

\begin{figure*}[h]
\hspace{-12mm}
\includegraphics[width=17.cm, height = 6.cm]{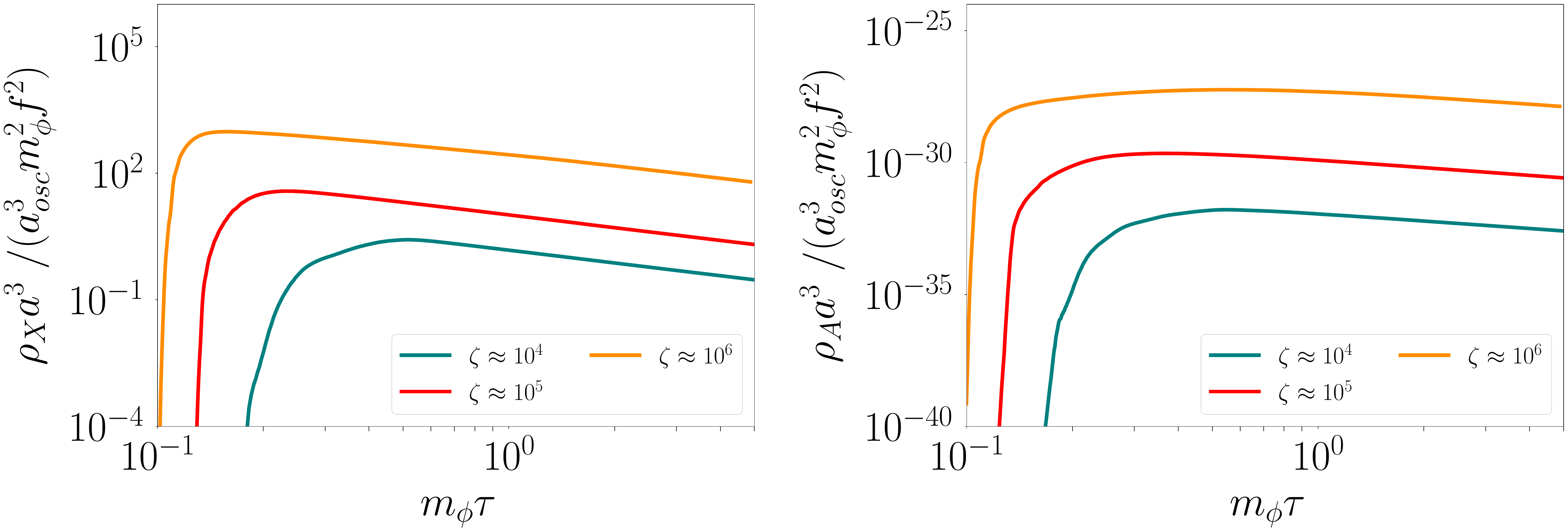}%
\caption{Electromagnetic energy densities obtained from simulations with $256^3$ grid points. $\rho_X$ and $\rho_A$ show a consistent behaviour with the results shown in Figure \ref{fig_energy_dens_String_f14}, which are obtained with $128^3$ grid points. The couplings are fixed to $g_{\phi X} = g_{\phi A X} = 80$. We also use $f = 10^{14}$ GeV and $m_{\phi} = 4.7 \times 10^{-8}$ eV, and assume that $T_R > f$.}
\label{fig_energy_dens_gauge_256}
\end{figure*}

The results presented in this paper are mostly obtained performing simulations with $128^3$ grid points. While in the literature post-inflationary axions are simulated on lattices with far higher resolutions \citep{Vaquero_2019, Buschmann_2020, OHare_2022, Pierobon_2023} (but without couplings to gauge fields), we check that the growth of $\rho_A$ displayed in Figures \ref{fig_energy_dens_String_f14} and \ref{fig_Pre_post_magnetogenesis} occurs also by increasing substantially the number of grid points, i.e. on grids with $256^3$ and $320^3$ points. Figure \ref{fig_energy_dens_gauge_256} shows indeed the same rapid growth of $\rho_X$ and $\rho_A$ in runs employing $256^3$ grid points, confirming the results presented in the main text (the same behaviour is observed in runs with $320^3$ grid points, not reported in this paper). We notice that the maximum value of $\rho_A$ in Figure \ref{fig_energy_dens_gauge_256} is higher than the values obtained in the main text. As mentioned in Appendix \ref{sec:appendix_A}, the reason behind such a difference is that the initialization procedure for the gauge fields depends, among other factors, on the grid spacing; in particular, runs with $256^3$ grid points have a higher initial $\rho_A$ with respect to runs with $128^3$ grid points.

\begin{figure*}[h]
\centering
\includegraphics[width=11.5cm, height = 8.5cm]{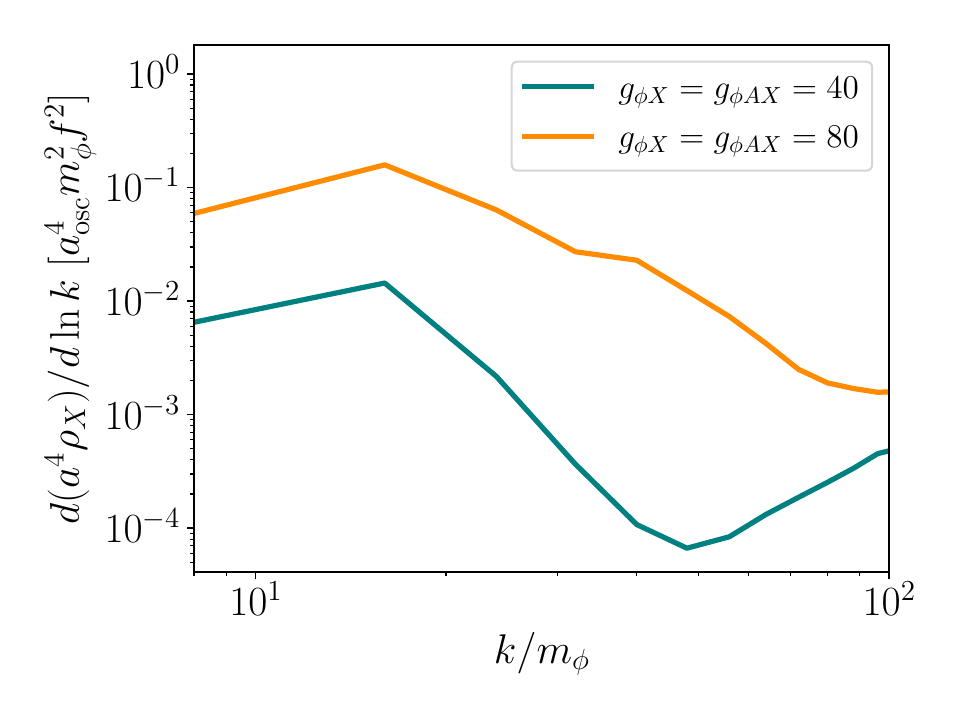}%
\caption{Spectrum of dark photons obtained from high resolution runs for two different sets of couplings and for $\zeta \approx 10^6$. The other parameters are fixed as in Figure \ref{fig_energy_dens_gauge_256}.}
\label{fig_Spectrum_string_dark_photon}
\end{figure*}

In Figure \ref{fig_Spectrum_string_dark_photon} we report the spectra of the sourced dark photons for $\zeta \approx 10^6$ and for different sets of couplings, which show a peak at relatively low momenta. We also note that the teal curve displays a power increase for $k/m_\phi \gtrsim 50$, which is probably an artefact due to the limited resolution available.

\newpage


%%%%%%%%%%%%%%%%%%%%%%%%%%%%%%%%%%%%%%%%%%
%	BIBLIOGRAPHY
%%%%%%%%%%%%%%%%%%%%%%%%%%%%%%%%%%%%%%%%%%

\label{Bibliography}

 \bibliographystyle{JHEP}
 \bibliography{biblio.bib}
\end{document}